\documentclass[journal]{IEEEtran}
\usepackage{cite}

\ifCLASSINFOpdf
\else
\fi

\usepackage{color}

\usepackage{lipsum}
\usepackage{url}
\usepackage{amsmath}
\usepackage{graphicx}
\graphicspath{{./figures/}}

\begin{document}
%
\title{Hybrid Open Points: an Efficient Tool for Increasing Network Capacity in Distribution Systems}


\author{Matthew~Deakin,~\IEEEmembership{Member,~IEEE,}
Ilias~Sarantakos,
David~M.~Greenwood,~\IEEEmembership{Member,~IEEE,}
Janusz~Bialek,~\IEEEmembership{Fellow,~IEEE,}
Phil~C.~Taylor,~\IEEEmembership{Senior Member,~IEEE,}
Wenlong~Ming,~\IEEEmembership{Member,~IEEE,}
and~Charalampos~Patsios,~\IEEEmembership{Member,~IEEE}
\thanks{M. Deakin, I. Sarantakos, D. M. Greenwood, J. Bialek, C. Patsios are with Newcastle University, Newcastle-upon-Tyne, UK. W. Ming is with Cardiff University. P. C. Taylor is with University of Bristol. Email: \texttt{matthew.deakin@newcastle.ac.uk}.}
}
%
%

\markboth{Accepted Article (IEEE PES Letters)}%
{Shell \MakeLowercase{\textit{et al.}}: Bare Demo of IEEEtran.cls for IEEE Journals}

%



\maketitle

\begin{abstract}
This letter introduces the Hybrid Open Point (HOP), a device consisting of an electromechanical switch connected in parallel with a power converter, for the purpose of providing additional network capacity in interconnected distribution systems. The HOP switch is used for bulk power transfer at low-cost, whilst the HOP converter provides targeted power transfer when the HOP switch is open. The device can replace either a Normally Open Point (Type 1 HOP) or a Normally Closed Point (Type 2 HOP). Simple interconnection and teed interconnection configurations are studied considering fault level and radiality constraints, with realistic use-cases identified for both HOP types. The HOP is shown to provide secure network capacity more cost-effectively than the classical Soft Open Point.
\end{abstract}


%
\IEEEpeerreviewmaketitle

\section{Introduction}
\IEEEPARstart{H}{ybrid} smart grid technologies are designed to augment the low-cost and efficient operation of AC systems with the flexibility of power electronic-based systems to improve network performance. A wide range of these `semiconductor-assisted' technologies have been proposed, including Hybrid On-Load Tap Changers \cite{Rogers2014low}, Hybrid Circuit Breakers \cite{he2019high}, Hybrid AC/DC Microgrids \cite{wei2020temporally}, Doubly-Fed Induction Generators \cite{abad2011doubly}, and more generally Hybrid AC-DC systems with a mix of AC and DC lines \cite{ahmed2019energy}.

The electrification of heat and transport in Net Zero systems will require Distribution System Operators (DSOs) to find new ways to meet secure network capacity requirements. If a substation is fed by two circuits without interconnection, the secure capacity is limited to just 50\% of the total rating of those two circuits. This utilisation can be increased by increasing the transfer capacity, provided with post-fault reconfiguration to provide interconnection between substations. Unfortunately, this does not always make the most use of all headroom available at an adjacent substation, as the demand that can be transferred may be limited by the available switching devices, layout of a given substation, or fault level considerations.

To increase this power transfer between substations, a Normally Open Point (NOP) can be replaced with a Soft Open Point (SOP), allowing for the radial operation of the network to be maintained whilst avoiding significantly increased fault level \cite{bloemink2010increasing,cao2016benefits,fuad2020soft,sarantakos2021reliability}. All of the additional headroom from adjacent substations can then be utilized, allowing the DSO to defer or even avoid network upgrades. It is worth noting, however, that the SOP has to carry not only the additional headroom, but also all of the load that the NOP would otherwise feed under post-fault conditions \cite{fuad2020soft}. This can mean that a large SOP may need to be installed for a relatively small increase in transfer capacity.

To address this issue, this letter proposes the Hybrid Open Point (HOP). The HOP design is presented in Section~\ref{s:hop_intro}, with Section~\ref{s:normal_sops} demonstrating how and when a HOP can provide additional secure network capacity more cost-effectively than a SOP (with 100\% utilization of the power converters). The lifetime benefits of HOPs are discussed in Section~\ref{s:benefits}, with salient conclusions closing the work in Section~\ref{s:conclusions}.

\section{The Hybrid Open Point}\label{s:hop_intro}

\begin{figure}\centering
\includegraphics[width=0.4\textwidth]{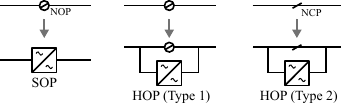}
\caption{Instead of replacing a Normally Open Point (NOP) with a Soft Open Point (SOP), the Hybrid Open Point (HOP) connects a power converter in parallel with a switch which may be normally open (Type 1) or normally closed (Type 2). When the HOP switch is closed, the switch carries the load; when the HOP switch is open, the HOP power converter can transfer power whilst maintaining radial operation of the network.}\label{f:hop_premise}
\end{figure}

The HOP consists of a switch in parallel with a power electronic converter, as illustrated in Fig.~\ref{f:hop_premise}. To distinguish between the configuration of the switch during normal operation, we further differentiate between a Type 1 HOP (the HOP switch is normally open) and a Type 2 HOP (the HOP switch is normally closed). The HOP switch is used to provide the bulk power transfer capacity of the integrated device, whilst the HOP power converter is used to transfer power across the device when the switch is open.

The HOP design is based on the observation that power converters are expensive per unit of continuous power rating compared to traditional AC technologies \cite{huber2017applicability}, with changes to the network configuration being particularly cost effective (where this is feasible) \cite{xiao2016tsc}. It is for this reason that it is so crucial for the HOP power converter to be placed in parallel with the HOP switch---the power transferred by the HOP converter is assumed to be just a fraction of the power transferred when the switch is closed. To ensure that costly network protection equipment does not need upgrading, the HOP converter would be controlled in a similar way to a SOP \cite{cao2016operating}, with the HOP switch in the same normal state as the NOP or NCP it is replacing.

\section{Using HOPs to Increase Secure Capacity:\\Simple and Teed Interconnection}\label{s:normal_sops}

In this section we study simple and teed interconnection between substations to demonstrate use cases of the HOP, and compare the required rating of the HOP power converter against a SOP providing equivalent functionality. Without loss of generality, we consider a DSO being mandated to operate securely under $N-1$ conditions (i.e., there should be no loss of load when one incoming HV circuit to a substation is out of service). We ignore trivial cases when there is already enough capacity to meet the demand, or when there is not enough capacity to meet all demand irrespective of topology. Voltage constraints are not considered, as it is assumed that traditional technologies are available to address these issues more cost effectively (e.g., voltage regulators). Following prior works, we use a nodal power balance approach, modelling the flow in feeders as the sum of downstream demands \cite{xiao2016tsc}.

\subsection{Simple Interconnection between Two Substations}

\begin{figure}\centering
\includegraphics[width=0.49\textwidth]{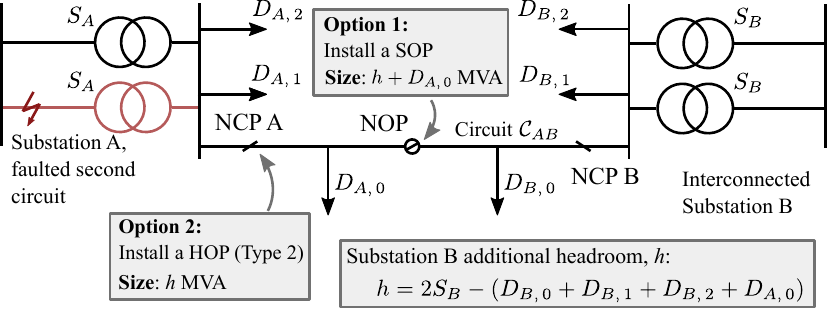}
\caption{Interconnection allows for increased secure capacity at Substation A, although fault level and radiality constraints may lead to latent headroom $h$ at Substation B being inaccessible with reconfiguration alone. This headroom can be utilised by replacing the NOP with a SOP of size $h + D_{A,\,0}$ (Option~1), or by replacing switch at Substation A, NCP A, with a Type 2 HOP of size $h$ (Option~2).}\label{f:hop_interconnection}
\end{figure}

Fig.~\ref{f:hop_interconnection} illustrates, at a high level, how two interconnected substations can increase their firm capacity by using automatic reconfiguration. When there is a fault on one of the circuits feeding Substation A, the NOP on circuit $\mathcal{C}_{AB}$ is closed and Normally Closed Point (NCP) A is opened, so the demand $D_{A,\,0}$ is transferred from Substation A to Substation B. 

If there is load growth at Substation A, so that the demand $D_{A,\,1}+D_{A,\,2}$ is greater than the circuit rating $S_{A}$, then additional capacity must be provided. We consider the case when additional headroom $h$ could be used to reduce net demand at Substation A. This additional headroom would be based on (for example) thermal constraints on circuit $\mathcal{C}_{AB}$; in this case we assume that Substation B's total capacity is the limiting factor.

Depending on the configuration of Substation A and the contribution of network components to fault level, there will be a limit to the amount of demand that can be transferred by reconfiguration alone. When this is the case, either a HOP or a SOP could be used to transfer the additional capacity, as follows.
\begin{itemize}
\item Option~1: replace the NOP on circuit $\mathcal{C}_{AB}$ with a SOP, sized to meet the feeder demand $D_{A,\,0}$ plus the additional headroom $h$. When a fault occurs on a circuit at Substation A, the SOP injects $D_{A,\,0}+h$ to meet the feeder demand $D_{A,\,0}$ and provide additional headroom $h$.
\item Option~2: install a Type 2 HOP in place of NCP A. The HOP power converter is sized to meet the additional headroom $h$ only. When a fault occurs on the incoming circuit at Substation A, the NOP closes, the HOP NCP opens, and the HOP power converter injects additional headroom $h$~MVA into Substation A.
\end{itemize}
In this case, both devices are providing the additional headroom $h$, with the HOP power converter of size $h$ whilst the SOP rating would be $h+D_{A,\,0}$, such that
\begin{equation}\label{e:hop_sop}
\mathrm{HOP/SOP\,power\,rating\,ratio}=\frac{h}{h+D_{A,\,0}}\,.
\end{equation}
For example, for headroom $h$ that is 25\% of the demand $D_{A,\,0}$ the HOP would only be 20\% the size of the equivalent SOP (although we note that a SOP could provide other operational benefits \cite{cao2016benefits}).

\subsubsection{Example}
The Haxby Road network \cite{sarantakos2019enhanced} has transformer size $S_{A}=15$~MVA, and is interconnected to Huntington New Lane, $S_{B}=24$~MVA. From 2030 to 2033, the Haxby Road demand could grow from 13.3~MVA to 16.7~MVA (the Net Zero Early scenario in \cite{npg2020dfes}), leading to a firm capacity shortfall of 1.7~MVA. Feeder B3 has load of $D_{A,\,0}=0.8$~MVA and is connected through a 5.5~MVA rated cable \cite{sarantakos2019enhanced} with load $D_{B,\,0}=0.6$~MVA. Huntington New Lane still has sufficient headroom $h>1.7$~MVA for $D_{A,\,0}$ to be transferred via reconfiguration, plus the additional 0.9 MVA required to meet the projected load growth. Therefore, as in Fig.~\ref{f:hop_dg_interconnection}, this load growth could be met either with a 1.7~MVA SOP installed at the end of Feeder B3 (Option 1), or an 0.9~MVA Type 2 HOP installed at the head of Feeder B3 (Option 2). The nodal power balance method (i.e., ignoring losses) was found to have less than 1\% relative error as compared to a full non-linear power flow solution, justifying its use in these calculations.

\subsection{Simple Interconnection with a Fault Level Constraint}

\begin{figure}\centering
\includegraphics[width=0.49\textwidth]{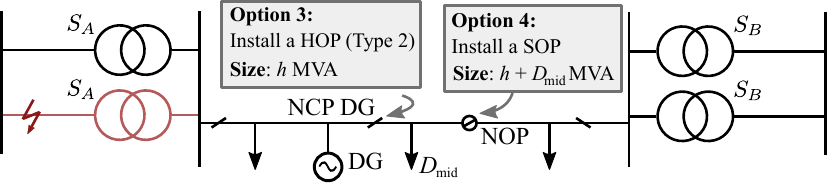}
\caption{If DG fault current limits the reconfiguration that can be done when there is a network outage, Options~3 and 4 both allow the additional headroom $h$ to be transferred, but Option~3 (using a Type 2 HOP) avoids oversizing the device by placing the HOP adjacent to the DG (at NCP DG).}\label{f:hop_dg_interconnection}
\end{figure}

Fig.~\ref{f:hop_dg_interconnection} illustrates a variation of the simple interconnected network, where the fault current of a DG limits reconfiguration (due to, e.g., the switchgear ratings at Substation B). In particular, it is assumed that the switch NCP~DG must be opened if the NOP is closed when there is a network outage. In this case, an $h$~MVA Type 2 HOP installed adjacent to the DG (Option~3) can transfer the same additional headroom $h$ as a $h+D_{\mathrm{mid}}$~MVA SOP installed in place of the NOP (Option~4). In this case, the HOP is only $h/(h+D_{\mathrm{mid}})$ the rating of the SOP.

\subsection{Teed Interconnection between Three Substations}
\begin{figure}\centering
\includegraphics[width=0.49\textwidth]{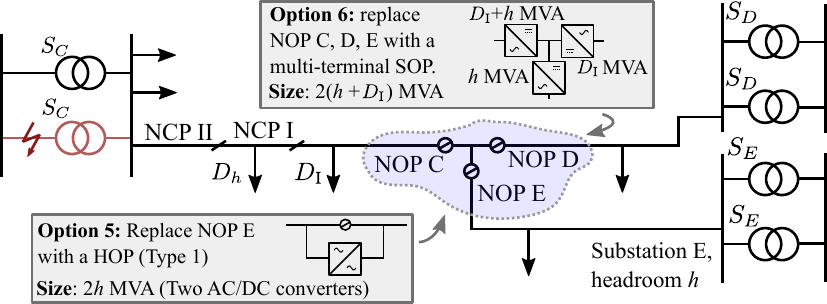}
\caption{In the case of a teed section of network, replacing NOP~E with a Type 1 HOP (Option~5) allows for the controlled injection of power into the circuit linking Substations C and D without having to install a fully rated multi-terminal SOP (Option~6).}\label{f:hop_teed}
\end{figure}

A teed interconnection case is shown in Fig.~\ref{f:hop_teed}. Prior to the installation of the HOP, under outage conditions NCP I would open, NOP~C and NOP~D would close, so that demand $ D_{\mathrm{I}} $ can be transferred to Substation D. Radial constraints at the teed section prevent any headroom $h$ at Substation~E being utilized. In this case, the replacement of NOP~E with a Type 1 HOP allows for additional load $D_{h}$ to be transferred during a network outage by the injection of $h$~MVA (Option 5), or using a suitably sized multi-terminal SOP (Option 6). For the multi-terminal SOP sized as illustrated in Fig.~\ref{f:hop_teed}, the ratio of power converter ratings of the HOP and SOP is $ h/(h+D_{\mathrm{I}}) $.

\subsection{Summary: Use Cases for a HOP}

As outlined at the start of this section, there needs to be an active security constraint and sufficient headroom at interconnected substations for a HOP to potentially provide additional firm capacity. In addition, all of the thermal ratings of any lines and cables would also need to be taken into account for calculating available headroom. Finally, there must be sufficient physical space at the substation (or along a feeder) for a device to be installed.

With those points in mind, the three cases envisioned for the HOP can be summarised as follows.
\begin{itemize}
\item When there are radiality constraints due to large lumped loads, a HOP can still be used to meet a fraction of that lumped load when there is a circuit outage. The lumped load could represent a large individual load, a radial spur, or adjacent feeders (e.g., demand $D_{A,\,1}$ in Fig.~\ref{f:hop_interconnection}) which cannot be transferred to an interconnected circuit (circuit $\mathcal{C}_{AB}$) due to the substation configuration.
\item When fault level constrains reconfiguration, the HOP can be installed at the network location where the protection becomes insufficient, irrespective of whether that point is normally open or normally closed. This constraint could be due to DG (Fig.~\ref{f:hop_dg_interconnection}) or insufficient fault current as distance from protection increases.
\item When there is a multi-terminal switching point (e.g., a teed section, as in Fig.~\ref{f:hop_teed}), a HOP can inject power to support the other feeders connected at that point. In these cases, there needs to be a configuration of the network so that the additional headroom can be utilised (e.g., NCP II can be opened in Fig.~\ref{f:hop_teed} to allow $D_{h}$ to be transferred).
\end{itemize}
A Type 1 HOP with the same rating as a SOP is identical, save for the additional HOP switch. Where a HOP that is installed is smaller than a SOP, there will be reduced opportunities for ancillary benefits \cite{cao2016benefits}, although this will only be a significant issue if the SOP is heavily dependent on these additional value streams for profitability.

\section{Benefits Analysis of a HOP}\label{s:benefits}
HOPs providing secure capacity can defer network reinforcement (or even avoid it entirely). Reinforcement deferral provides a net benefit to society due to the time value of money--the present value of the equivalent saving for a delay of $N$ years with discount rate $d$ can be calculated as
\begin{equation}\label{e:tvom}
\mathrm{Cost\,Reduction}\,(\%) = 100 - \dfrac{100}{(1 + d)^{N}}\,.
\end{equation}
For example, a 5-year reinforcement deferral with a 3.25\% discount rate is equivalent to a 14.8\% cost reduction. HOPs could therefore provide a cost-effective way of managing incrementally increasing demand.

The HOP can also provide ongoing operational benefits. A marginal benefit of $B$ \$/yr over a device lifetime of $N$ years has an lifetime operational benefit (present value) of
\begin{equation}\label{e:ops_benefit}
\mathrm{Lifetime\,Operational\,Benefit}\,(\$) = B \sum_{i=1}^{N}\dfrac{1}{(1 + d)^{i}}\,.
\end{equation}
For example, if a 3 MVA HOP reduces the losses in a network on average by 44 kW (as for the SOP in the central case of \cite{cao2016benefits}), the annual loss reduction is 385 MWh/yr; if the wholesale energy price over this period is \$50/MWh, then over ten years the present value of the benefit would be \$162,320.

To determine the socially optimal investment approach, DSOs should carry out a cost-benefit analysis, comparing the HOP against all other credible solutions over the project lifetime. Specific further benefits that a HOP could provide, either through reinforcement deferral \eqref{e:tvom} or via operational benefits \eqref{e:ops_benefit} include the following:
\begin{itemize}
\item The HOP power converter can provide reactive power support/power factor correction.
\item With the choice of a suitable power converter technology, the HOP could provide power conditioning, reducing (for example) network unbalance or harmonics.
\item A Type 1 HOP embedded within the network provides all of the benefits of an equivalently sized SOP at that location \cite{cao2016benefits}.
\end{itemize}

\section{Conclusions}\label{s:conclusions}
This letter compares the proposed Hybrid Open Point against the conventional Soft Open Point, particularly in the context of the provision of network capacity in interconnected systems. As a HOP can replace both normally closed and normally open points, it provides additional flexibility for DSO planning engineers compared to a SOP, allowing for smaller, more cost-effective devices with higher utilisation. Future work on HOPs should focus on developing suitable mathematical methods to determine optimal placement and operation of these devices in the context of flexible loads and DGs alongside more sophisticated network security standards.

AC systems are low-cost, with mature voltage transformation and protection technologies. Conversely, DC systems based on power electronics can flexibly control power flows to fully utilise existing assets. We conclude that only by embracing Hybrid systems can the full potential of both of these paradigms be realised to cost-effectively reach Net Zero.


%

%

\section*{Acknowledgment}

The authors would like to thank Dr Jialiang Yi of UK Power Networks and Anuj Chhettri of Northern Powergrid for their helpful discussions. This work was funded by ESPRC Grant No. EP/S00078X/1 (Supergen Energy Networks Hub 2018) and EPSRC Grant no. EP/T021969/1 (Multi-energy Control of Cyber-Physical Urban Energy Systems).

\ifCLASSOPTIONcaptionsoff
  \newpage
\fi

\bibliography{refs_sops_configuration}{}
\bibliographystyle{IEEEtran}

\end{document}